\documentclass[twocolumn,reprint,superscriptaddress,amsmath,amssymb,aps,prb,longbibliography]{revtex4-2}
\usepackage[colorlinks=true, linkcolor=blue, citecolor=blue,urlcolor=blue]{hyperref}
\usepackage{amsfonts}
\usepackage{subfigure}
\usepackage{amsmath}
\usepackage{mathrsfs}
\usepackage{amssymb}
\usepackage{amsbsy}
\usepackage{epsfig}
\usepackage{graphicx}
\usepackage{epstopdf}
\usepackage{mathdots}
\usepackage{color}
\usepackage{cleveref}
\usepackage{float}
\usepackage{graphicx}
\usepackage{rotating}
\usepackage{physics}
\usepackage{nicefrac}
\usepackage{braket}
\usepackage{physics}
\usepackage{xcolor}

\begin{document}
\title{Biorthogonal dynamical quantum phase transitions in non-Hermitian topological superconductors}

\author{Haoran Gu}
\affiliation{Department of Physics, Jiangsu University, Zhenjiang, 212013, China}

\author{Yubo Zhao}
\affiliation{Department of Applied Mathematics, Xi’an Jiaotong-Liverpool University, Suzhou, 215123, China}

\author{Siyuan Cheng}
\affiliation{Department of Physics, Jiangsu University, Zhenjiang, 212013, China}

\author{Yuee Xie} \altaffiliation{yueexie@ujs.edu.cn}
\affiliation{Department of Physics, Jiangsu University, Zhenjiang, 212013, China}  

\author{Xiaosen Yang} \altaffiliation{yangxs@ujs.edu.cn} 
\affiliation{Department of Physics, Jiangsu University, Zhenjiang, 212013, China}  

\author{Yuanping Chen} \altaffiliation{chenyp@ujs.edu.cn}
\affiliation{Department of Physics, Jiangsu University, Zhenjiang, 212013, China}  

\date{\today}

\begin{abstract}
Dynamical quantum phase transitions in non-Hermitian systems pose fundamental challenges due to the intrinsic biorthogonality of their eigenstates. In this work, we extend a biorthogonal framework to investigate dynamical quantum phase transitions in non-Hermitian topological superconductors. Taking the non-Hermitian Kitaev chain as a prototypical model, we construct an associated-state formalism and reformulate the Loschmidt rate function, dynamical topological order parameter, and dynamical Fisher zeros. Within this framework, we find that the critical times at which dynamical quantum phase transitions occur differ from those based on the conventional self-normal approaches. We further analyze momentum-resolved subsystems at critical momenta and demonstrate the robustness of the biorthogonal framework. Our work highlights the essential role of biorthogonality in nonequilibrium dynamics and establishes a consistent theoretical framework for dynamical quantum phase transitions in non-Hermitian topological superconductors.

\end{abstract}

\maketitle
\section{Introduction}
In recent years, non-Hermitian physics has attracted significant attention due to phenomena absent in Hermitian systems \cite{lee_anomalous_2016,leykam_edge_2017,yaoEdgeStatesTopological2018,gong_topological_2018,el-ganainy_nonhermitian_2018,lee_topological_2019,song_nonhermitian_2019,yokomizoNonBlochBandTheory2019,kawabataSymmetryTopologyNonHermitian2019,scheibner_nonhermitian_2020,bergholtzExceptionalTopologyNonHermitian2021,li_nonhermitian_2021,jing_biorthogonal_2024,reisenbauer_nonhermitian_2024}, such as the exceptional points \cite{zhong_winding_2018,zhu_simultaneous_2018,liu_dynamically_2021,krol_annihilation_2022,guoExceptionalNonAbelianTopology2023,klauckCrossingExceptionalPoints2025}, and the non-Hermitian skin effect (NHSE) \cite{longhiProbingNonHermitianSkin2019,songNonHermitianSkinEffect2019,liCriticalNonHermitianSkin2020,zhangCorrespondenceWindingNumbers2020,liang_dynamic_2022,zhang_universal_2022,kawabata_entanglement_2023}. Unlike Hermitian Hamiltonians with real spectra and orthogonal eigenstates, non-Hermitian systems generally exhibit complex eigenvalues and require distinct right and left eigenstates \cite{jones-smith_nonhermitian_2010,martinezalvarez_nonhermitian_2018,sim_observables_2025}, which satisfy a biorthogonal normalization condition rather than the conventional orthogonality relation \cite{kunst_biorthogonal_2018,edvardsson_phase_2020,sun_biorthogonal_2021}. This biorthogonal structure fundamentally modifies the description of quantum states and observables and leads to an unconventional spectral topology \cite{miriExceptionalPointsOptics2019,dingNonHermitianTopologyExceptionalpoint2022,lvCurvingSpaceNonHermiticity2022,zhou_entanglement_2022,zhouObservationGeometrydependentSkin2023}. These distinctive spectral and biorthogonal properties also enrich the notion of topological phases in non-Hermitian systems, allowing for novel topological phase transitions \cite{shen_topological_2018,
okuma_topological_2020}. Notably, non-Hermiticity can induce a breakdown of the conventional bulk-boundary correspondence \cite{helbigGeneralizedBulkBoundary2020,xiaoNonHermitianBulkBoundary2020,yangNonHermitianBulkBoundaryCorrespondence2020,zirnsteinBulkBoundaryCorrespondenceNonHermitian2021,zhangObservationNonHermitianTopology2021,chen_characterizing_2022,yokomizoNonBlochBandsTwodimensional2023,nakamuraBulkBoundaryCorrespondencePointGap2024}.

Dynamical quantum phase transitions (DQPTs) characterize dynamical critical behavior in nonequilibrium quantum systems and have been widely studied over the past decades \cite{heyl_dynamical_2014,budich_dynamical_2016,kennes_controlling_2018,wang_simulating_2019,yang_observation_2019,tian_observation_2020}. It was first introduced in the Bose-Hubbard Model\cite{schutzhold_sweeping_2006,fischer_bogoliubov_2008} and Ising model \cite{heyl_dynamical_2013,kriel_dynamical_2014,bhattacharjee_dynamical_2018,guo_observation_2019,chen_experimentally_2020}, and later extended to Floquet systems \cite{zamani_floquet_2020,jafari_floquet_2022,naji_engineering_2022,naji_dissipative_2022}, finite temperature \cite{abeling_quantum_2016,mera_dynamical_2018,mondal_finitetemperature_2023}, and mixed states \cite{bhattacharya_mixed_2017,heyl_dynamical_2017,lang_dynamical_2018}. The signatures of DQPTs have also been observed in recent experiments \cite{bernien_probing_2017,nie_experimental_2020,zhou_nonhermitian_2021,hamazaki_exceptional_2021,mueller_quantum_2023}. Its extension to non-Hermitian systems remains an open and actively developing problem \cite{nehra_anomalous_2024,dora_work_2024,fu_anatomy_2025,zhang_selfnormal_2025}. Existing approaches to non-Hermitian DQPTs often rely on self-normal bases and normalization procedures \cite{zhou_dynamical_2018}, which can lead to ambiguous or even unphysical results. This limitation calls for a consistent formulation of transition probabilities within a proper biorthogonal framework \cite{edvardsson_nonhermitian_2019,yoshimura_nonhermitian_2020,xu_dynamical_2021,lin_experimental_2022,geier_nonhermitian_2022,zhai_nonequilibrium_2022,Decoding_zhao_2026}. Recently, a systematic theory based on biorthogonal bases has been developed for the non-Hermitian Su–Schrieffer–Heeger model \cite{jing_biorthogonal_2024}, highlighting the essential role of biorthogonality in identifying genuine dynamical criticality. Nevertheless, its extension to non-Hermitian topological superconducting systems remains largely unexplored.

In this work, we investigate DQPTs in non-Hermitian superconductors within the biorthogonal framework. The special biorthogonality is not merely a formal modification of the inner-product structure, but has direct and substantial consequences for nonequilibrium dynamical behavior. To consistently characterize these effects, we construct an associated-state formalism and reformulate the central quantities used to diagnose DQPTs, including the Loschmidt rate, the dynamical topological order parameter (DTOP), and the dynamical Fisher zeros, entirely within the biorthogonal framework. On this basis, we demonstrate that the critical times extracted from the biorthogonal Loschmidt rate are systematically shifted relative to those obtained within conventional self-normal approaches, revealing a qualitative difference between the two descriptions. We further analyze the momentum-resolved subsystems at the critical momenta and show that the resulting dynamical criticality remains robust under the biorthogonal formulation. These results not only clarify the decisive role played by biorthogonality in non-Hermitian nonequilibrium dynamics, but also establish a consistent theoretical framework to study DQPTs in non-Hermitian topological superconductors.

\section{Model and dynamical quantum phase transition}
We consider the non-Hermitian Kitaev chain with complex chemical potentials in momentum space \cite{kitaev_unpaired_2001,zeng_nonhermitian_2016}, described by the Hamiltonian $H = \sum_k \psi_{k}^\dagger \mathcal{H}_{k} \psi_{k}$ in the basis  $\psi_{k} = (c_{k}, c_{-k}^\dagger)^{T}$, with
\begin{align}
\mathcal{H}_{k} = \Delta \sin k \sigma_y - (t_1\cos k + \mu)\sigma_z ,
\label{Hamk}
\end{align}
where $t_1$, $\Delta$, and $\mu$ denote the nearest-neighbor hopping amplitude, $p$-wave pairing amplitude, and complex chemical potential $\mu= \mu_r + i \gamma$, where $\mu_r$ and $\gamma$ are the real and imaginary parts of the chemical potential, respectively. $\boldsymbol{\sigma}$ represents the Pauli matrices.  The eigenvalues are given by $\epsilon_{k} = \pm \sqrt{\Delta^2 \sin ^2 k + ( t_1\cos k + \mu)^2 }$, and the corresponding eigenstates are $\ket{u_{k\pm}}$. The phase boundaries are obtained from the gap-closing condition $\epsilon_{k} = 0$, i.e., $\mu_r = -t_1\cos k$ and $\gamma^2 = \Delta^2\sin^2 k$. In the $\mu_r - \gamma$ parameter plane, the condition can be written as
\begin{align}
    \left( \frac{\mu_r}{t_1} \right)^2 + \left( \frac{\gamma}{\Delta} \right)^2 = 1.
\label{phaseboundary}
\end{align}

In non-Hermitian systems, the conventional overlap $\left|\langle \psi(t)\vert \psi\rangle\right|^2$ between an initial state $|\psi\rangle$ and its time-evolved state $|\psi(t)\rangle$, as used in Hermitian systems, does not admit a direct probabilistic interpretation since the eigenstates are no longer normalized. To avoid such ambiguities, we employ the associated-state construction and define a biorthogonal inner product that yields a natural normalization for time-dependent observables \cite{jing_biorthogonal_2024}. For an arbitrary state $\ket{\psi_k}=\sum_k c_k\ket{u_k}$, we define its associated state $\ket{\tilde{\psi}_k}$ by replacing each basis vector with the corresponding biorthogonal basis vector \cite{jing_biorthogonal_2024}:
\begin{align}
\ket{\psi_k} = \sum_{k} c_{k} \ket{u_{k}} 
\quad \leftrightarrow \quad
| \tilde{\psi}_{k} \rangle = \sum_{k} c_{k} \ket{\tilde{u}_{k}},
\label{associatedstate}
\end{align}
where $c_{k} = \langle \tilde{u}_k \ket{\psi_k}$. In practical calculations, we work with the Hermitian conjugate of the associated state, $\langle \tilde\psi_k | = \sum_{k} c_k^\ast \langle \tilde{u}_k |$, and define the inner product between $\ket{\psi}=\sum_k c_k\ket{u_k}$ and $\ket{\phi}=\sum_m d_m\ket{u_m}$ by
\begin{align}
\langle \phi,\psi\rangle
\equiv \langle \tilde{\phi}|\psi\rangle
&=\sum_{m,k}\langle \tilde{u}_m|\,d_m^{*}c_k\,|u_k\rangle
=\sum_k d_k^{*}c_k.
\label{innerproduct}
\end{align}
Accordingly, the norm of the state $\ket{\psi_k}$ is given by $\sqrt{\braket{\tilde{\psi}_k|\psi_k}}$. With these definitions, the transition probability between $\ket{\psi}$ and $\ket{\phi}$ in a non-Hermitian system can be written as
\begin{align}
p=\frac{\langle \tilde{\psi}\!\mid\!\phi\rangle\langle \tilde{\phi}\!\mid\!\psi\rangle}
{\langle \tilde{\psi}\!\mid\!\psi\rangle\langle \tilde{\phi}\!\mid\!\phi\rangle},
\label{transitionprobability}
\end{align}
where the denominator provides a natural normalizing factor. Under this definition, the transition probability lies in $[0,1]$ and retains a clear probabilistic interpretation.

The static aspects of non-Hermitian systems are now comparatively well understood, but their dynamical aspects are still under debate. Though the equation $\ket{\Psi(t)} = e^{-iH \cdot t}\ket{\Psi(0)}$ describes how an arbitrary initial state $\ket{\Psi(0)}$ evolves under a Hamiltonian $H$, $\ket{\tilde{\Psi}(t)} = e^{-iH^\dagger \cdot t}\ket{\tilde{\Psi}(0)}$ \cite{tang_dynamical_2022} can lead to complex transition probabilities when $H \neq H^\dagger$. To address this issue, we introduce the associated state $|\tilde{\Psi}(t)\rangle = \sum_k c_k |\tilde{u}_k\rangle$, with $c_{k} = \langle \tilde{u}_k\!\mid\!\Psi(t) \rangle$. The overlap between $|{\Psi}(0)\rangle$ and $|{\Psi}(t)\rangle$ is then given by
\begin{align}
\mathcal{L}(t)=
\frac{\langle \tilde{\Psi}(0)\!\mid\!\Psi(t)\rangle \langle \tilde{\Psi}(t)\!\mid\!\Psi(0)\rangle}
{\langle \tilde{\Psi}(t)\!\mid\!\Psi(t)\rangle \langle \tilde{\Psi}(0)\!\mid\!\Psi(0)\rangle},
\label{L(t)}
\end{align}
which can be regarded as the biorthogonal Loschmidt echo in non-Hermitian systems.

We examine a quench process described by $H = H_k^f \theta(t) + H_k^i \theta(-t)$. At $t=0^{-}$, the system is characterized by $H_k^i$, while at $t=0^{+}$, it is governed by $H_k^f$. Therefore, the initial state $ |\Psi(0)\rangle = \otimes_k |u_{k-}^i\rangle$ subsequently evolves under the post-quench Hamiltonian $H_k^{f}$. To diagnose DQPTs, the biorthogonal Loschmidt echo can be written as $\mathcal{L}(t)=\prod_{k} g_k(t)$, with 
\begin{equation}
g_k(t)=\frac{|\cos\small(\epsilon_{k}^ft\small) - i\sin\small(\epsilon_{k}^ft\small) \langle{\tilde{u}_{k-}^{i}}|{\frac{H_{k}^{f}}{\epsilon_{k}^{f}}|}{u_{k-}^{i}\rangle}|^{2}}{\left\langle \tilde{u}_{k-}^{i}(t) \middle| u_{k-}^{i}(t) \right\rangle},
\label{gkt}
\end{equation}
where $\ket{u_{k-}^{i}(t)} = e^{-i H_{k}^{f} \cdot t} \ket{u_{k-}^{i}}$. Here, $g_k(t)$ represents the contribution from each momentum mode. As in the Hermitian case, the biorthogonal DQPTs occur when $\mathcal{L}(t_c) = 0$ \cite{zhou_dynamical_2018}. In the thermodynamic limit, the biorthogonal Loschmidt rate is defined as \cite{jing_biorthogonal_2024}
\begin{align}
\operatorname{LR}(t) = -\lim_{N \to \infty} \frac{1}{N} \ln \mathcal{L}(t),
\label{LRt}
\end{align}
where $N$ is the size of the system. Whenever $\mathcal{L}(t) = 0$, $\operatorname{LR}(t)$ becomes nonanalytic, typically exhibiting cusp-shaped singularities.

The criterion for the occurrence of a DQPT is $\mathcal{L}(t) = 0$, and the solution is given by
\begin{align}
t_{n}(k)=\frac{\pi}{2\epsilon_{k}^{f}}(2n+1) - \frac{i}{\epsilon_{k}^{f}} \tanh^{-1} \mel{\tilde{u}_{k-}^{i}}{\frac{H_{k}^{f}}{\epsilon_{k}^{f}}}{u_{k-}^{i}},
\label{tnk}
\end{align}
with integer $n$. A DQPT occurs when there exists a critical momentum $k_c$ such that a curve of the dynamical Fisher zeros $Z_n(k) = it_n(k)$ crosses the imaginary-time axis, thereby producing a critical time $t_c$. In finite-size systems, the discreteness of the momenta causes the temporal profile of $\operatorname{LR}(t)$ to exhibit a cusp rather than a divergence. Therefore, the analysis of the Fisher-zero trajectories in the complex-time plane provides a more robust characterization of the critical structure.

To characterize the topological aspects of non-Hermitian DQPTs, we introduce the biorthogonal DTOP \cite{vajna_topological_2015,budich_dynamical_2016}
\begin{align}
\nu(t)=\frac{1}{2\pi}\int_{0}^{2\pi} dk\partial_{k}\phi_{k}^{\mathrm{G}}(t).
\label{DTOP}
\end{align}
Here, the geometric phase is defined as $\phi_k^{G}(t)=\phi_k(t)-\phi_k^{\rm dyn}(t)$, where $\phi_k(t)$ is the total phase of $\langle \tilde{u}^{i}_{k}|u^{i}_{k}(t)\rangle$, and the biorthogonal dynamical phase is given by \cite{gong_piecewise_2018}
\begin{align}
\phi_{k}^{\mathrm{dyn}}(t) = & -\int_{0}^{t} ds\frac{\langle \tilde{u}_{k-}^{i}(s) | H_{k}^{f} | u_{k-}^{i}(s) \rangle
}{\langle \tilde{u}_{k-}^{i}(s) | u_{k-}^{i}(s) \rangle
} \nonumber\\
& +\frac{{i}}{2} \ln \langle \tilde{u}_{k-}^{i}(t) | u_{k-}^{i}(t) \rangle .
\label{dynamicalphase}
\end{align}
When the eigenvalue is $\pi$-periodic in $k$, the upper limit of the integral in Eq.~\eqref{DTOP} can be taken as $\pi$ instead of $2\pi$ \cite{mondal_anomaly_2022}. 

\begin{figure}
\includegraphics[width=0.45\textwidth]{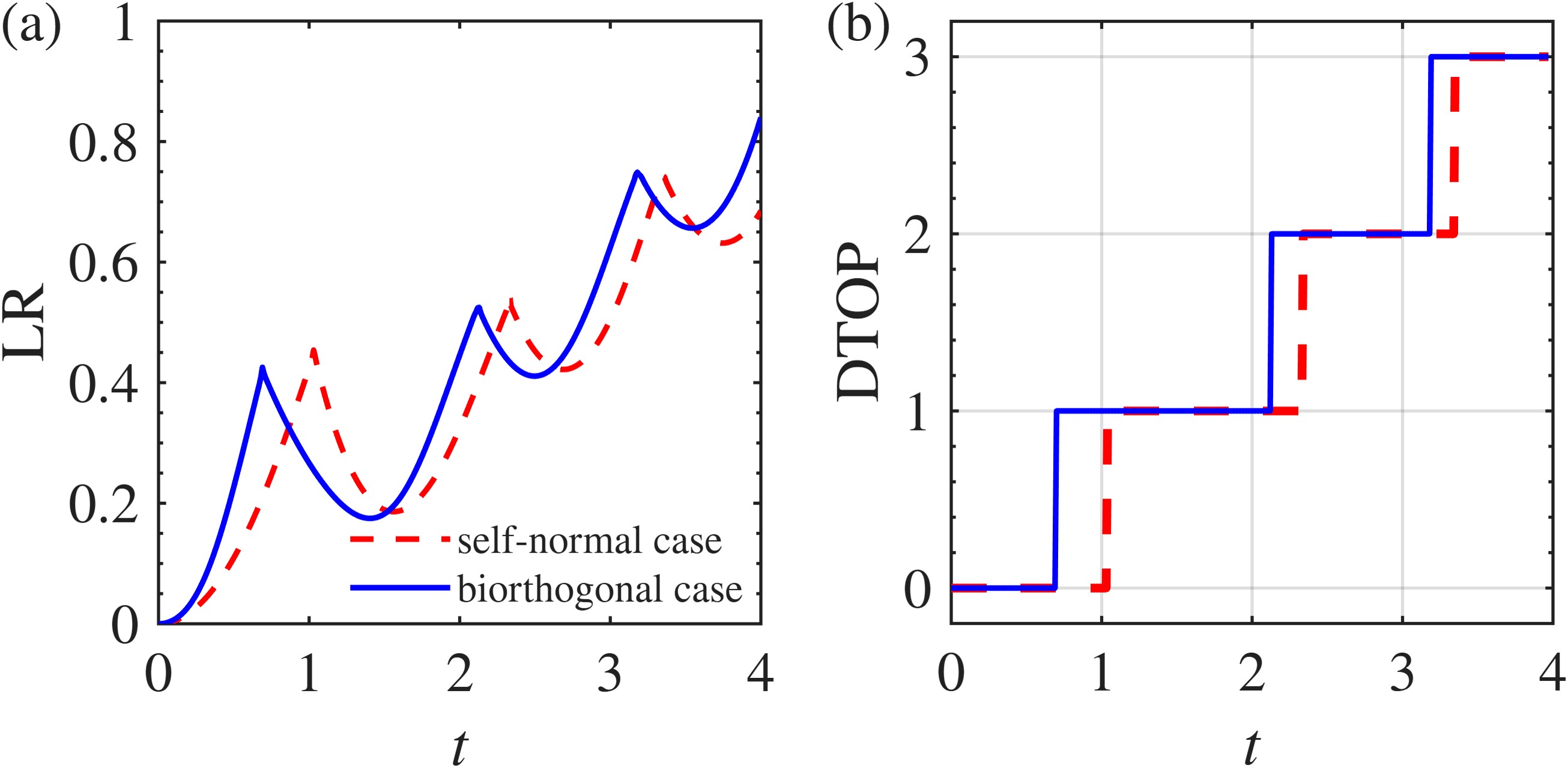}
\caption{(a) Nonanalytic behavior of the biorthogonal Loschmidt rate obtained from Eq.~\eqref{LRt}, clearly visible at critical times. (b) Biorthogonal DTOP evaluated from Eq.~\eqref{DTOP}, showing a unit jump at the critical time in agreement with the Loschmidt rate. Common parameters: $t_1 = 1$, $\Delta = 0.9$, $\gamma = 0.5$, $\mu_r^i = 0.25$, and $\mu_r^f = 1.7$.}
\label{fig1} 
\end{figure}

In order to further study DQPTs, we also evaluate the transition probability between $\ket{u^{i}_{k-}(t)}$ and $\ket{u^{i}_{k+}}$ \cite{jing_biorthogonal_2024}:
\begin{equation}
p(k,t)=
\frac{\langle \tilde{u}^{i}_{k+}|u^{i}_{k-}(t)\rangle\,
\langle \tilde{u}^{i}_{k-}(t)|u^{i}_{k+}\rangle}
{\langle \tilde{u}^{i}_{k-}(t)|u^{i}_{k-}(t)\rangle}.
\label{pkt}
\end{equation}
Whenever there exists a critical momentum $k_c$ satisfying $p(k_c,t) = 1$ at a certain time $t$, the system undergoes a DQPT, and the momentum $k_c$ coincides with the critical mode identified from the zeros of $\mathcal{L}(t)$. By expanding the time-evolved state in terms of the initial eigenstates as $\ket{u^{i}_{k-}(t)}=c_1\ket{u^{i}_{k-}}+c_2\ket{u^{i}_{k+}}$, where $c_1=\langle \tilde{u}^{i}_{k-}|u^{i}_{k-}(t)\rangle$ and $c_2=\langle \tilde{u}^{i}_{k+}|u^{i}_{k-}(t)\rangle$, the transition probability between $\ket{u^{i}_{k-}(t)}$ and $\ket{u^{i}_{k+}}$ can be rewritten as $p(k,t)=c_2^{*}c_2/(c_1^{*}c_1+c_2^{*}c_2)$.

\section{Comparison between the conventional self-normal and the biorthogonal cases}

\begin{figure}
\includegraphics[width=0.45\textwidth]{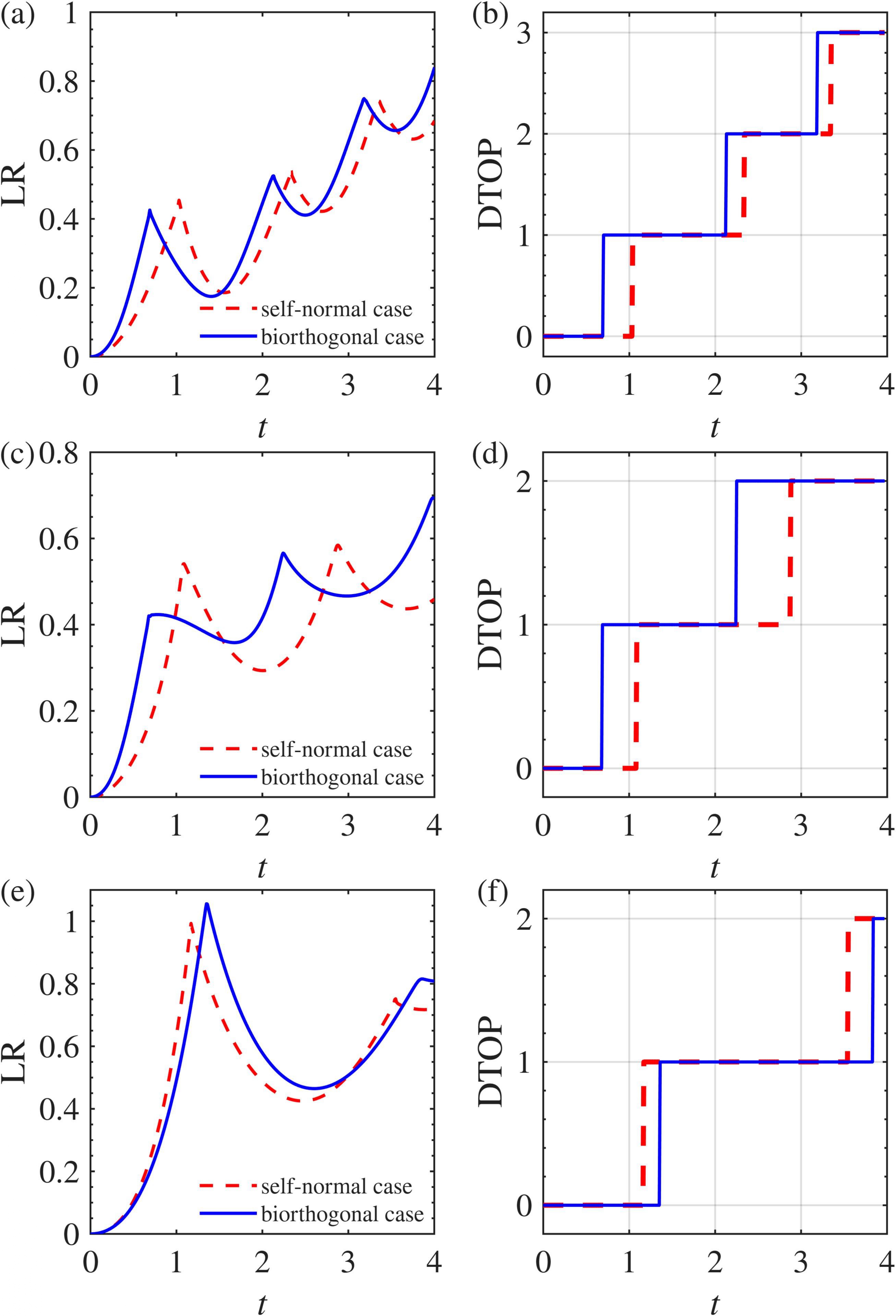}
\caption{DQPTs for three different quench processes (from top to bottom): a quench from the topological phase to the topologically trivial phase with ($\mu_r^i$, $\mu_r^f$) = ($0.3$, $1.7$), a quench inside the trivial phase with ($\mu_r^i$, $\mu_r^f$) = ($-0.6$, $0.7$), and a quench from the topologically trivial phase to the topological phase with ($\mu_r^i$, $\mu_r^f$) = ($1.7$, $-0.3$). (a), (c), and (e) show the Loschmidt rate, while (b), (d), and (f) show the corresponding DTOP. Common parameters: $t_1=1$, $\Delta=0.9$, and $\gamma=0.5$.}
\label{fig2} 
\end{figure}

For comparison with the biorthogonal Loschmidt echo, Fig.~\ref{fig1}(a) also shows the conventional Loschmidt echo $\mathcal{L}(t) = |\langle\Psi(0)|\Psi(t)\rangle|^2$, together with an enforced normalization factor $\bigl|e^{-iH_k^f \cdot t}|u_{k-}^i\rangle\bigr|^2$ \cite{zhou_dynamical_2018}. We consider a quench from the topological phase to the topologically trivial phase and plot the self-normal and biorthogonal Loschmidt rates. The critical times extracted from the two schemes do not coincide: those obtained from the biorthogonal construction occur systematically earlier. A similar discrepancy has also been reported for equilibrium quantum phase transitions in non-Hermitian systems, where the critical points inferred from the biorthogonal and self-normal fidelity generally are generally distinct. Importantly, owing to the intrinsic biorthogonality of eigenstates, the biorthogonal fidelity yields a physically consistent critical point in the thermodynamic limit. For the same reason, nonequilibrium quantum phase transitions are naturally characterized by the corresponding time-dependent biorthogonal fidelity, i.e., the biorthogonal Loschmidt echo \cite{sun_biorthogonal_2021}. Figure~\ref{fig1}(b) displays a corresponding feature: the DTOP changes by an integer whenever a DQPT occurs, thereby providing a topological characterization of the dynamical criticality.

\begin{figure}
\includegraphics[width=0.45\textwidth]{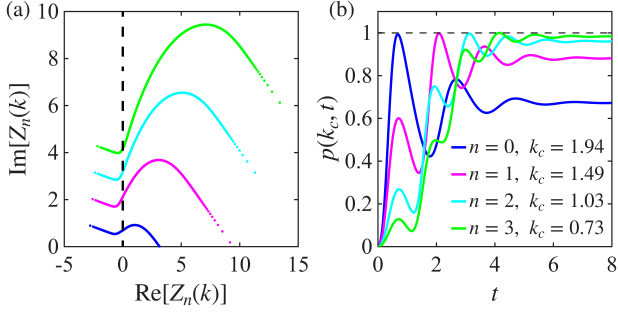}
\caption{(a) Dynamical Fisher zeros $Z_n(k)$ for the representative quench process in Fig.~\ref{fig1}, shown for $n = 0$, $1$, $2$, $3$. (b) Corresponding transition probability between $\ket{u_{k_c-}^{i}(t)}$ and $\ket{u_{k_c+}^{i}}$ at the critical momentum.}
\label{fig3} 
\end{figure}

We analyze two additional quench processes by varying $\mu_r$ while keeping $\gamma$ fixed, to demonstrate that this behavior is generic rather than specific to a particular quench process. Figure~\ref{fig2} summarizes three representative quench processes. Although, for one of the three quench processes, the critical time extracted from the biorthogonal formulation occurs later, the biorthogonal approach consistently identifies the correct critical times in all three cases. These critical times also coincide with the corresponding topological signatures, as manifested by the DTOP jumps shown in Fig.~\ref{fig2}.

To further elucidate the critical structure in the complex-time plane, we investigate the dynamical Fisher zeros $Z_n(k)=i\,t_n(k)$ \cite{yang_statistical_1952,lee_statistical_1952,fisher_theory_1967,andraschko_dynamical_2014,peng_experimental_2015,zauner-stauber_probing_2017}. Since a physical DQPT can only occur at real times, a geometric criterion naturally follows: when a curve of $Z_n(k)$ intersects the imaginary axis, which means $\mathrm{Re}[Z_n(k)]=0$, the intersection corresponds to a real critical time $t_{c}$, and the associated momentum is the critical momentum $k_c$. Thus, the crossings of the imaginary axis provide a direct criterion for extracting critical momentum and critical time. 

We also evaluate the transition probability $p(k_c,t)$ at the critical momenta. Owing to the biorthogonality of eigenstates, the self-normal scheme cannot reproduce $p(k_c,t)=1$. It follows that $p(k_c,t)=1$ implies $\langle \tilde{u}^{i}_{k_c+}|u^{i}_{k_c-}(t)\rangle = 1$, and hence $\langle \tilde{u}^{i}_{k_c-}|u^{i}_{k_c-}(t)\rangle = 0$. Since the Loschmidt echo can equivalently be expressed as
\begin{equation}
\mathcal{L}(t) =\prod_k
\frac{\langle\tilde{u}_{k-}^i|u_{k-}^i(t)\rangle\,
\langle\tilde{u}_{k-}^i(t)|u_{k-}^i\rangle}
{\langle\tilde{u}_{k-}^i(t)|u_{k-}^i(t)\rangle}.
\label{Ltuki}
\end{equation}
Figure~\ref{fig3}(b) indicates that multiple critical momenta $k_c$ can exist, each associated with a distinct critical time $t_{c}$. Moreover, a characteristic pattern is observed: for $t<t_{c}$, $p(k_c,t)$ exhibits $n$ local maxima and reaches $1$ at the $(n+1)$-th maximum, whereas for $t\gg t_{c}$, it tends to a steady state.

By employing a biorthogonal framework, we study DQPTs in a non-Hermitian Kitaev chain. This formulation applies to general non-Hermitian Hamiltonians with complex spectra, and we illustrate its implementation and physical consequences together with a comparison to the self-normal treatment.

\begin{figure}
\includegraphics[width=0.45\textwidth]{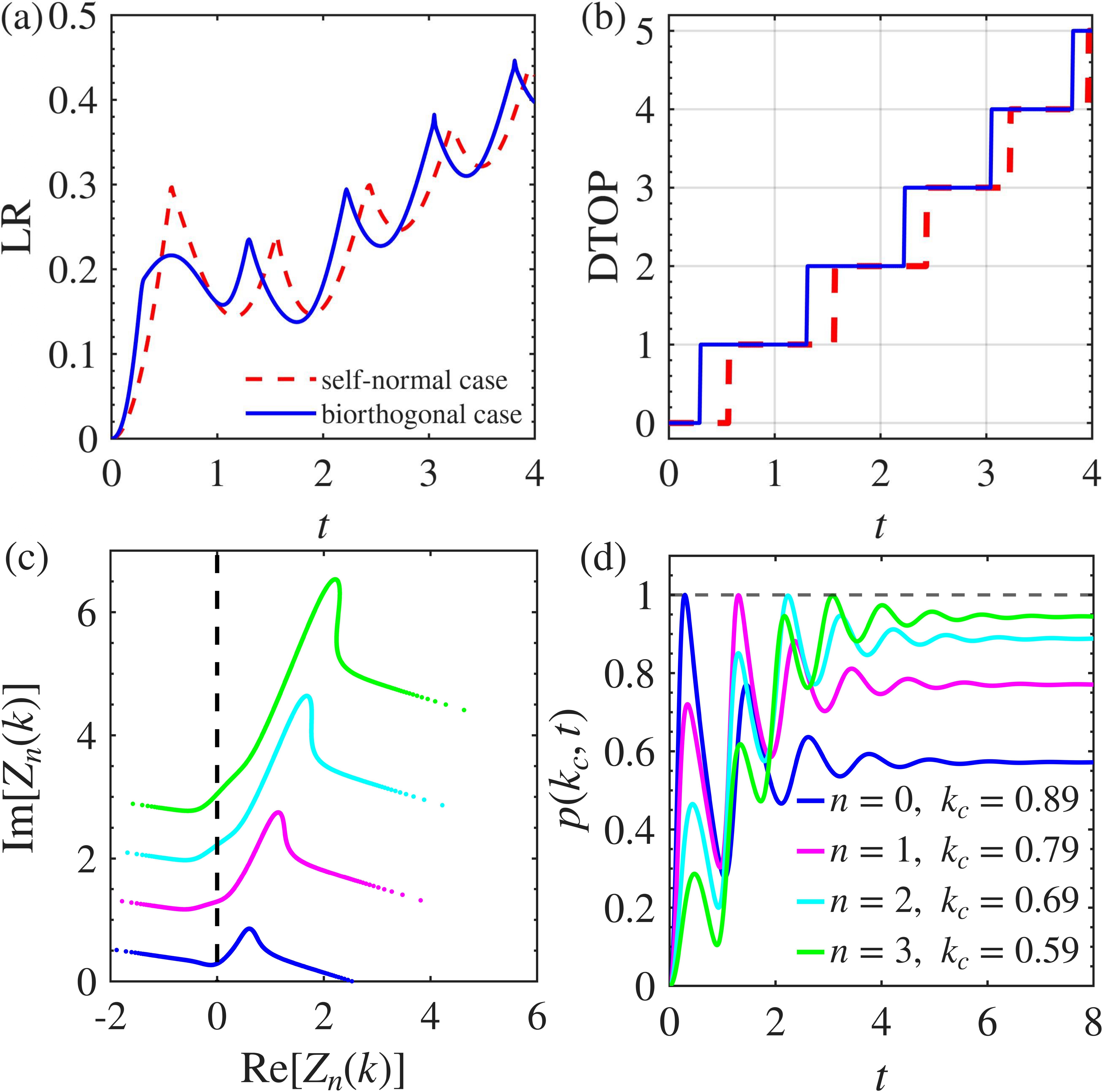}
\caption{(a) Nonanalytic behavior exhibiting the critical times. (b) DTOP jumps corresponding to the evolution of $\operatorname{LR}(t)$. (c) Curves of $Z_n(k)$ crossing the imaginary axis and displaying DQPTs for $n = 0$, $1$, $2$, $3$. (d) Corresponding transition probability between $\ket{u_{k_c-}^{i}(t)}$ and $\ket{u_{k_c+}^{i}}$ at the critical momentum. Common parameters: $t_1 = 1$, $t_2 = 0.7$, $\Delta = 0.9$, $\gamma = 0.5$, $\mu_r^i = -0.5$, and $\mu_r^f = 2.2$.}
\label{fig4}
\end{figure}

\section{Extension to the Case with Next-Nearest-Neighbor Hopping}
Based on the non-Hermitian Kitaev chain discussed above, we consider an extension by incorporating a next-nearest-neighbor hopping amplitude \cite{wang_spontaneous_2015}, described by $H = \sum_k \psi_{k}^\dagger \mathcal{H}_{k} \psi_{k}$ in the basis  $\psi_{k} = (c_{k}, c_{-k}^\dagger)^{T}$, with
\begin{align}
\mathcal{H}_{k}
&=\Delta \sin k\, \sigma_y-\left[t_1\cos k+t_2\cos(2k)+\mu\right]\sigma_z.
\label{HamkNNN}
\end{align}
Here, $t_2$ denotes the next-nearest-neighbor hopping amplitude, while all other parameters remain unchanged. The phase boundaries are determined by $\mu_r = -t_1\cos k -t_2\cos(2k) $ and $\gamma^2 = \Delta^2\sin^2 k$, which can be written as
\begin{align}
\left(\mu_r+t_2-2t_2\frac{\gamma^{2}}{\Delta^{2}}\right)^{2}
=t_1^{2}\left(1-\frac{\gamma^{2}}{\Delta^{2}}\right).
\label{NNNpb}
\end{align}

For this extended model, we investigate the Loschmidt rate, the DTOP, the dynamical Fisher zeros, and $p(k_c,t)$ within the biorthogonal framework. For a representative quench process from a topological phase to a topologically trivial phase, the critical times predicted by the self-normal and biorthogonal Loschmidt rate remain systematically shifted. The jumps of the DTOP are consistent with the geometric criterion based on the dynamical Fisher zeros, indicating that the curves of $Z_n(k)$ cross the imaginary axis. Moreover, $p(k_c,t)$ satisfies $p = 1$ at the critical time, corresponding to the nonanalytic behavior of $\operatorname{LR}(t)$. These observations demonstrate that the proposed biorthogonal dynamical framework remains applicable to general non-Hermitian Hamiltonians with complex spectra.

\section{conclusion and discussion}
In summary, we adopt a biorthogonal framework to investigate DQPTs in the non-Hermitian Kitaev chain with complex chemical potentials. By introducing the concept of the associated state, we redefine the transition probability between an arbitrary state and its time-evolved state, thereby removing the negative values that arise in the self-normal case through a proper incorporation of the intrinsic biorthogonality. We further investigate the biorthogonal DTOP and the dynamical Fisher zeros, both of which consistently verify the critical times associated with DQPTs. In addition, the analysis of the momentum-resolved subsystem at the critical momenta provides further support for the physical consistency of the biorthogonal description and clarifies. We also extend our analysis to a non-Hermitian Kitaev chain with next-nearest-neighbor hopping and find that the main dynamical features remain qualitatively unchanged, indicating that the biorthogonal framework is not restricted to the simplest nearest-neighbor model. More broadly, the biorthogonal framework developed here can be generalized to spinful non-Hermitian superconductors with $s$-wave and $d$-wave pairing. Our results therefore demonstrate the robustness of the biorthogonal description and highlight the importance of a biorthogonal formulation for characterizing nonequilibrium critical phenomena in non-Hermitian topological superconductors. An important implication of the present work is that the biorthogonal formulation may also be necessary when connecting non-Hermitian DQPTs to experiments. Since the conventional self-normal treatment can shift the critical times, an experimental interpretation based solely on self-normal overlaps may become ambiguous in genuinely non-Hermitian settings. By contrast, the biorthogonal framework adopted here yields a consistent transition probability and provides compatible diagnostics through the Loschmidt rate, the DTOP, the Fisher-zero trajectories, and the condition $p(k_c,t) = 1$. This internal consistency suggests that the biorthogonal description can serve as a more reliable theoretical basis for analyzing future experiments on non-Hermitian quench dynamics. 

\section{Acknowledgements}
This work was supported by National Natural Science Foundation of China (Grant Nos. 12174157, 12174158, 12272378), Natural Science Foundation of Jiangsu Province (Grant No. BK20231320), Open Research Project of the National Key Laboratory of Infrared Science and Technology(SITP-SKLIP-YB-2025-09).

\bibliographystyle{apsrev4-1}
\bibliography{DQPTreference}

\end{document}